\documentclass[aps,showpacs,preprintnumbers,amsmath,amssymb, 
eqsecnum,
twocolumn, tightenlines
]{revtex4}

\usepackage{graphicx}

\newcommand{\be}{\begin{eqnarray}}
\newcommand{\ee}{\end{eqnarray}}

 \newcommand{\gsim}{\mathrel{\hbox{\rlap{\lower.55ex \hbox {$\sim$}}
                   \kern-.3em \raise.4ex \hbox{$>$}}}}
\newcommand{\lsim}{\mathrel{\hbox{\rlap{\lower.55ex \hbox {$\sim$}}
                   \kern-.3em \raise.4ex \hbox{$<$}}}}

\def\roughly#1{\mathrel{\raise.3ex\hbox{$#1$\kern-.75em%
\lower1ex\hbox{$\sim$}}}}
\def\lsim{\roughly<}
\def\gsim{\roughly>}

\begin{document}


\title{ Comments on the CMS discovery of the ``Ridge" \\
in High Multiplicity pp collisions at LHC  }
\author { Edward Shuryak}
\address { Department of Physics and Astronomy, State University of New York,
Stony Brook, NY 11794}
\date{\today}

\begin{abstract}
A very recent paper by the CMS collaboration \cite{cms_ridge} has created large discussion in the media, which call it important but did not explain why, in some
places even calling it  ``unundestandable". While it is of course too soon to know what causes the correlations in question, a very similar observation
in heavy ion collisions at RHIC has rather simple explanation related to explosion of high energy density matter. Perhaps this observation is the first
hint for an explosive behavior in pp, which was anticipated and looked for for decades, yet never  been seen. 
\end{abstract}
\maketitle
\section{The Ridge}
Very recent paper by the CMS collaboration \cite{cms_ridge} has created large discussion in the field, in LHC community and even in the media. The experimentalists themselves, who have found this effect, certainly can tell a lot about the ideas which drived them, test/comparisons with  various Monte-Carlo generators etc.
However, for obvious reasons,  in the paper itself and in CMS CERN presentation they
prefered not to discuss the basic physics  but simply keep to the pure stated facts.  Thus discussion  in the media, blogs etc
were commenting  on its potential importance without any  explanations of what those may possibly be.  
Since it has created active discussion in our group as well, I  wrote up those comments, for wider public use. No part of their content  is new: most
of it is well known in the heavy ion community. Only some historic remarks are original.

Techically, the experiment study the correlations of two charge particles in specially selected high multiplicity events of pp collisions at LHC.
The finding is that in such case, unlike in the usual (or ``minimally biased" as they as called) collisions, one finds correlation between particles which is very wide in rapidity 
difference $\Delta \eta$ and yet concentrated at small azimuthal angle  $\delta \phi < 1$. 

While such correlations have never been seen in pp collisions, a similar phenomenon has been known in AuAu collisions at RHIC. While strickly speaking
even in the latter case we do not yet have a completely established explanation, we are quite sure that it has to do with explosion of new form of matter,
the Quark-Gluon Plasma (QGP) which is by now very well documented and studied. Whether the newly found ridge in pp has a similar origin or not
we do not yet know: by the end of these comments we will discuss further possible tests which are to be performed to answer this question.

 \begin{figure}[t!]
\includegraphics[width=1.7in]{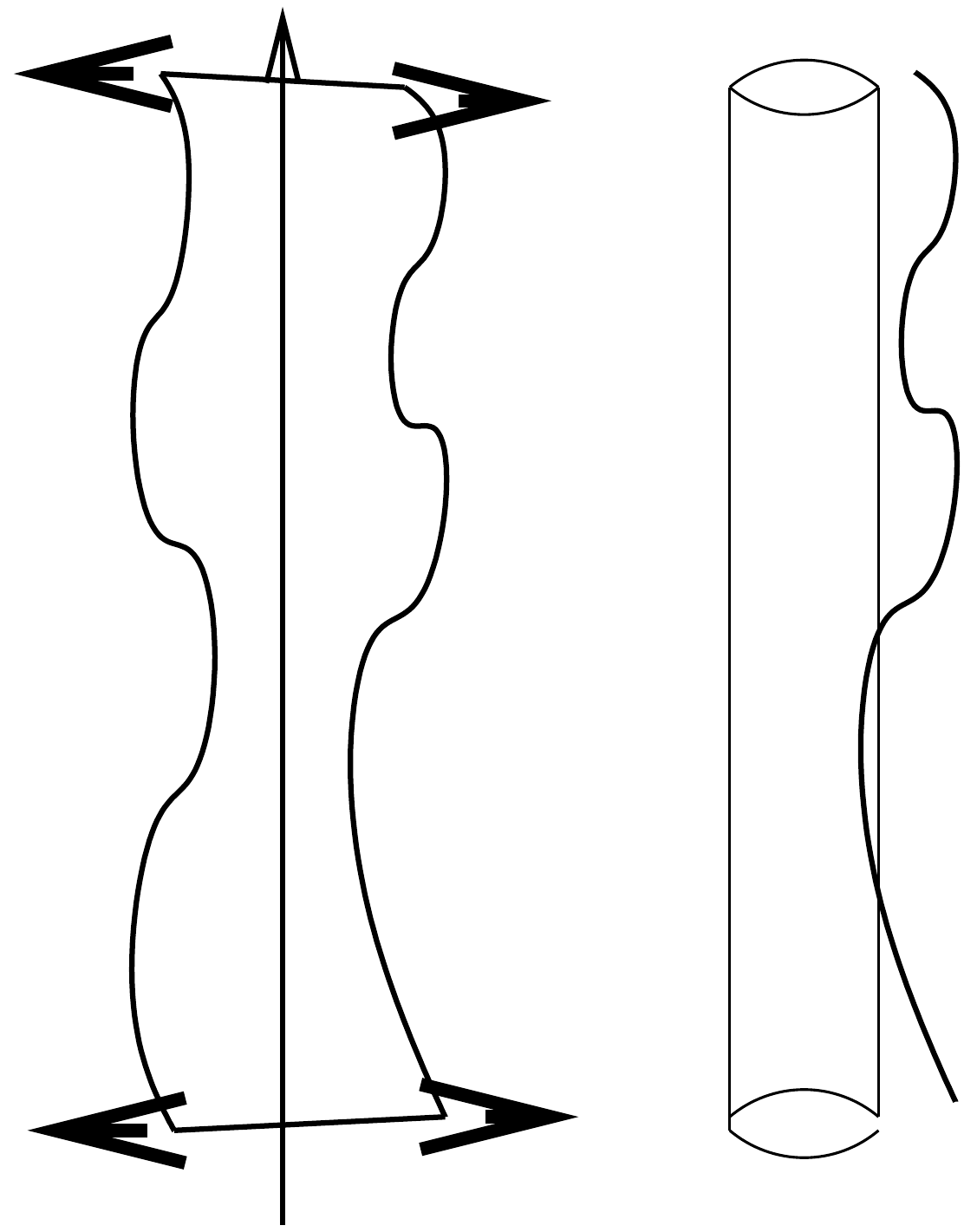}
\caption{ (left) Two strings, stretched in the beam direction (vertical line with an arrow) which also move away from each other. (right) a string placed near the stick of explosive}
\end{figure}
 
\section{What does the ridge mean?}
A very general view of pp collisions is that during very short time in which two protons pass each other, they still can exchange some color charges,
as a result of which the departing systems remain connected to each other by ``color strings" containing flux of color-electric fields. As they are
longitudinally stretched, they break into pieces (called clusters) which then decay into finally observed particles which are sufficiently stable to fly into detectors.
These clusters mover relativistically relative to each other, and cannot possibly exchange any information, except at the very moment when the strings are produced.

Since breaking happens as a sequence of independent quantum unrelated fluctuations, it is natural that clusters get kicked into random directions, and also decay
isotropically (in their frame) independently of each other. So
any information (e.g. about a direction of any one of the final particles)
are quite soon forgotten, if one goes along the string (which is called rapidity direction).
 And indeed, the usual (normal multiplicity) pp events show only short-range correlations in rapidity (related with the longitudinal coordinate along the beam,
along which the strings are being stretched). The decays are isotropic around the beam, thus no special angles $\phi$ can be selected. Summary of the usual string correlations: small $\Delta\eta$, arbitrary $\Delta\phi$.

The ``ridge" is the {\bf opposite} correlation: it is present in  a narrow strip of azimuthal angle $\Delta\phi\sim 1 \ll 2\pi$ and very wide in rapidity (large $\Delta\eta$).   

So, the first question to ask is how  some long and rapidly stretched object (string) can consistently decay into a rather narrow range of angles
in azimuth? What may cause all the secondaries be directed in the same direction, taking into account that places where all the  decays happen move relativistically away from each other, preventing any communication between them by causally? 

A possibility is that this happens  because the string moves in this direction, {\bf as a whole}. 

Taking this as a working hypothesis, let us first ask {\bf where to} (or where from) this vector  may point?

The question looks very obvious at first sign: by definition zero azimuth points along the momentum of the first particle. So,  may it simply mean
that when a (mini)jet is emitted, the emission is asymmetric in the reaction plane? For example, it may be  because there are strings connecting
the jet to some forward/backward moving remnants of the proton with matching (anti)color. Indeed, there are such effects known, they have been
observed in e+e- and DIS. However, those are small,  included in some event generators, which however all failed to show any ``ridge" in simulations. 

Since we discuss some small probability fluctuation, can it be that we have spontaneous motion of two partons in {\bf both} protons
moving away from each other? It will then create two strings, rapidly moving away from each other in the transverse plane (Fig.1 left), and generating
two strings, decaying into at $\phi\approx 0$ and  $\phi\approx \pi$ directions , as observed? 

What is wrong with this (and the previous) options is that they both
 ignore the main experimental fact known about the "ridge": It {\em does not exist in the usual pp events}, and  
only appears {\bf  if the multiplicity is unusually high}.   So, why is  high  multiplicity  needed ? 
    
Here is another simple option. The angular direction in the transverse plane is $not$ related with the decay directions, which are still uncorrelated and randomly directed, in the 
string rest frame. The correlation appears because of the string's   {\bf position}, 
relative to a {\bf stick of explosive} (Fig.1 , right) placed nearby. Whatever the string decay products are, they  will be blown away after the
explosive went off, in the same direction {\bf away from
 the  stick's center}. Perhaps the explosion is only happening if the multiplicity is high enough.

 The ridge observed in the AuAu collisions at RHIC is most likely is of that nature. In this case (unlike the pp events at LHC) we know the size of the explosive
 (the radius of Au nucleus), the equation of state of matter produce. We know that spectra of all secondaries are
 naturally explained by a hydrodynamical radial flow resulting from the pressure gradient.  We know 
 to which velocity a string at its surface should be accelerated  (about $v_\perp \sim 0.7c$), which roughly explains the width $\delta\phi$ of the correlation.
 By changing centrality of the collision we can change this velocity, and we see that the width changes accordingly.  We also know in which $p_t$ window the
 hydro effect should be seen, and this matches the observation of the ridge as well.
 
  So, the open question is:  can pp events also generate an explosion, strong enough to kick the 
remnants of the string decay into relativisitc motion in transverse direction? We know that the ``usual" (mean multiplicity) pp events are too dilute for that, but
perhaps the high multiplicity events selected by CMS in their highest bin $N_{ch}> 110$ can. 
Maybe it will also be hydrodynamical blast. If not, perhaps  it will still make a ``wind" of particles,  strong enough to explain the data.

\section{Bits of history}
In a very well known story, Enrico Fermi have calibrated the first nuclear explosion with small pieces of paper, carried by a shock wave to a distance he used as his input.
For a great physicist like him, it was not even nacessary to look at the blast itself or use all other data collected. What I am saying is that the CMS ridge is a bit
like Fermi's pieces of paper.  (In a less known story Fermi was the first, in 1951, to discuss statistical equilibration in high energy pp collisions.)

Thirty-plus years ago, a brand new pp collider,  ISR , also at CERN, had started its operation, and when the first identified particle spectra appeared 
I (and my then student Zhirov) have been busy checking if they show any sign of transverse expansion. The answer unfortunately has been {\bf negative} \cite{Shuryak:1979ds}.
But what is important is how we approach the question. If the flow has certain velocity, and it carries particles of different mass -- pions, kaons, nucleons, deuterons -- 
the momenta they will have $p=Mv/\sqrt{1-v^2}$ would all be different. It will create departure from the usually expected ``transverse mass"  $m_\perp^2=p_\perp^2+m^2 $ scaling,
$$ dN/ dp_\perp^2 \sim f(m_\perp)  $$ 
normally expected from string decays. But the data stubbornly show no evidences of $m_t$ scaling violation: we end up speculating why possible expansion is not happening.

About 15 years ago Bjorken and collaborators start working on MiniMax experiment at Fermilab Tevatron. One of the ideas they tried has been triggering on high multiplicity events. 
Perhapse here, Bjorken argued, the system will gets dense enough, and at the same time small and rapidly expanding, so that it will produce his favorite object,
a Disoriented Chiral Condensate (DCC). Unfortunately, it did not happen   \cite{Brooks:1999xy}. Was there at least some kind of  explosion, a deviation from independent
minijet/string fragmentations? Maybe: as far as I know  it was never seriously studied.

At about the same time, fixed target CERN SPS heavy ion data have finally shown unambiguous signs of the radial flow, which also was in crude agreement
with hydrodynamics \cite{Hung:1997du} . The elliptic flow however still was not working properly. Perhaps, we thought, it may work at RHIC, providing our predictions \cite{Teaney:2000cw,Kolb:1999es}.
And indeed it all worked marvelously, without any adjustement, see \cite{Teaney:2001av,Kolb:2001qz}.  New additions to the story are the third harmonics 
in azimuth \cite{Alver:2010gr}
and the fourth \cite{Lacey:2010fe}, which together further constrain the viscosity and provide better understanding of the limits of hydrodynamics.
What we know about QGP suggest that  the ``mean free path" (derived from viscosity) is unexpectedly small, only  about .2 $fm$. So perhaps a proton is
macroscopic enough, and can hydrodynamically explode provided there is  enough matter produced inside it?

What was truly surprising was that hydro works for AuAu for about 99\% of the particle, up to $p_\perp \sim 2\, GeV$, and ends around 3 GeV due to jets. 
We have witness that the spectra of the pions and proton/antiprotons
cross each other, so that there are more antiprotons than pions in certain $p_t$ window. In fact baryon/meson ratio is increased by about factor 10, before it
drops at high $p_t$ to a small value expected from jet fragmentation. 

When the ridge has been discovered in Au Au collisions  it has been interpreted as a hydro flow phenomenon, as we discussed above.  
Additional support for it  was observation of identified  particle  spectra inside the ridge: those are quite similar to bulk spectra, with enhanced baryon/meson ratio.
They are slightly harder, because hydro at the edge has a bit larger velocity: it also fits well into the picture.
The best window for its observation is $p_t=1-3 \, GeV$, exactly the same as for the new CMS ridge. Is it a coincidence, or hydro may again be that good?

Let me finish with some unexpected theoretical result, related to transition to ``equilibration". In AdS/CFT framework ``thermal" or ``non-thermal" regimes have
 well defined meaning: those are collisions in which the black hole is or is not formed. Since we so far used it in the large-$N_c$ approximation, the 
 question is about classical gravity, and thus have simply yes or no answer. With a surprise Lin and myself  \cite{Lin:2009pn}
 have recently found that for example as a function of the impact parameter the switch between the two regimes happens sharply, with a jump.
 And experimentally, the entropy/participant is about the same for all ``thermal" RHIC collisions (of fixed energy) but it is different from that in pp and dAu one
 (see Fig.5 in the last Ref.), so there is kind of a jump there. When I was speaking about it at the seminar in NYU a year ago, Glannys Farrar asked if such thing may happen in pp, e.g. at LHC? ``Not in minimally bias collisions, I answered, but perhaps at sufficiently high multiplicity events." 
   
For clarity: I do not want to claim anything here:
the possibility to have explosion in high multiplicity pp has been widely anticipated in the community, see e.g. refs in [1] such as Wiedemann-Casalderrey suggestion to  look for its second angular harmonic \cite{CasalderreySolana:2009uk}. 

Everybody knew it is going to happen at some density, yet nobody knew when the multiplicity would be high enough. That is why
the current hints from CMS are so exciting: right or wrong, we will know after some further scrutiny.

\section{What to do next?}
After the last section, it is obvious. In fact the situation at the moment is quite unusual: subtle 2-particle correlations are measured by CMS before the
1-particle spectra! With about 0.3 million high-multiplicity
CMS events on tape, it should be doable: just more time is need to get particle ID and detector systems to be calibrated. (Remember, it is a maiden voyage of large
and very complicated ship, with brand new team and leadership!)

The first test is to see if there is any explosion. For this one has to plot the pion, kaon and nucleon spectra, and to see if they are consistent with
common  radial flow or not. At LHC one may wander also what heavier particles like $D,B$ and even $\psi$ mesons do: are they also blown away
to higher momenta?

More challenging task is to estimate the baryon/meson ratio inside {\em the ridge itself}, for $p_T=1-2 \, GeV$. If a high value would be found, it will  exclude any
jet-based  explanations right away. 

There is also a remaining homework for RHIC experiments as well, namely to study  the onset of the explosive regime. Recent low energy scan should take care of the energy dependence, but there are remaining issues related with
 understanding the transition to equilibration/hydro regime as a function of the system {\bf size}. While very peripheral collisions have large geometric (or Glauber)
 fluctuations, their account plus hydro reproduces elliptic flow data reasonably well, till the ``almonds" get just a couple fm wide. So,  we actually know that hydro does work
 for systems whose size is only factor 2 or 3 away
 from the size of the single nucleon!

\end{document}